\documentclass
[notitlepage,12pt,groupedaddress,secnumarabic,prd,showpacs,showkeys,onecolumn,nofootinbib]{revtex4}%
\usepackage{amsfonts}
\usepackage{amsmath}
\usepackage{amssymb}
\usepackage{graphicx}
\usepackage{appendix}
\usepackage{color}%
\begin{document}
\title{Small Deformations of Kinks and Walls}
\author{J.R. Morris}
\affiliation{Physics Dept., Indiana University Northwest, 3400 Broadway, Gary, Indiana
46408, USA}

\begin{abstract}
A Rayleigh-Schr\"{o}dinger type of perturbation scheme is employed to study
weak self-interacting scalar potential perturbations occurring in scalar field
models describing 1D domain kinks and 3D domain walls. The solutions for the
unperturbed defects are modified by the perturbing potentials. An illustration
is provided by adding a cubic potential to the familiar quartic kink potential
and solving for the first order correction to the kink solution, using a
\textquotedblleft slab approximation\textquotedblright. A result is the
appearance of an asymmetric scalar potential with different, nondegenerate,
vacuum values and the subsequent formation of vacuum bubbles.

\end{abstract}

\maketitle

\section{Introduction}

\ \ Exact solutions describing 1D domain kinks and 3D domain walls for
$\varphi^{4}$ scalar field theory with a symmetric potential of the form
$V_{0}(\varphi)=\frac{1}{4}\lambda(\varphi^{2}-a^{2})^{2}$ are well known,
with vacuum values located by $\varphi=\pm a$, and static solutions assume the
form $\varphi(x)=\pm a\tanh(kx)=\pm a\tanh(x/w)$, where $w$ is a
\textquotedblleft width parameter\textquotedblright\ for the kink/wall
\cite{Vilenkin}-\cite{KTbook}. However, the addition of a small perturbing
potential $V_{1}(\varphi)$ will, in general, distort the simple $\tanh(kx)$
solutions in some way that depends upon the form of $V_{1}(\varphi)$
\cite{Deform}-\cite{approx}.

\bigskip

\ \ An effort here is made to focus upon a Rayleigh-Schr\"{o}dinger type of
perturbation scheme resulting in corrections to the unperturbed solutions, the
corrections being due to the perturbing potential $V_{1}(\varphi)$. This
method involves an expansion of the solution $\varphi(x)$ in terms of powers
of an expansion parameter $g$, along with an expansion of the full potential
$V(\varphi)=V_{0}(\varphi)+V_{1}(\varphi)$ about the zeroth order solution
$\varphi_{0}(x)$ which solves the unperturbed equation of motion.

\bigskip

\ \ This perturbation scheme differs from the excellent one introduced by
Almeida, Bazeia, Losano, and Menezes \cite{Defect13} for $(1+1)$ dimensional
topological defects, wherein the unperturbed action $S_{0}(\varphi)$ is
supplemented by an additional perturbing action $S_{1}(\varphi)=\alpha\int
d^{2}xF(\varphi,X)$, where $\alpha$ is a very small parameter controlling the
perturbative expansion, $X=\frac{1}{2}\partial^{\mu}\varphi\partial_{\mu
}\varphi$, and $F(\varphi,X)$ is, in principle, an arbitrary function of
$\varphi$ and $X$ describing the perturbation to a kink-like defect in $(1+1)$
dimensions. Additionally, an example of the perturbation method that is
presented here differs from the examples of those provided in \cite{Defect13},
where the perturbations in \cite{Defect13} leave the total potential symmetric
about $\varphi=0$, with degenerate vacuum states.

\bigskip

\ \ The scheme posed here is illustrated with an example where the perturbing
potential is chosen to be $V_{1}(\varphi)=\frac{1}{3}\mu\varphi^{3}$, with
$\mu$ being a small mass parameter. An approximate solution for the first
order correction is obtained for this potential, with the use of a
\textquotedblleft\textit{slab approximation}\textquotedblright\ for the
unperturbed kink/wall defect \cite{JM95}. (For a very thin domain wall
described by a scalar field $\varphi$ with an associated energy scale
$E\sim\sqrt{\lambda}a\sim1$ GeV, the wall thickness is $\lesssim1$ fermi.) The
additional contribution $V_{1}(\varphi)$ causes the total potential
$V(\varphi)$ to become \textit{asymmetric}, with two slightly different vacuum
states $\varphi_{+}$ and $\varphi_{-}$, and two different, and
\textit{nondegenerate}, associated vacuum values $V^{+}(\varphi_{+})$ and
$V^{-}(\varphi_{-})$, so that $\Delta V=V^{+}-V^{-}\neq0$ for $\mu\neq0$.
Stress-energy components for the two different sides of the wall are
calculated, with $T_{\mu\nu}^{+}\neq T_{\mu\nu}^{-}$, indicating an
instability against bending, ending in a formation of a network of vacuum
bubbles. Without an efficient stabilizing mechanism, the bubbles subsequently
collapse, releasing radiation in the form of $\varphi$ boson particles of mass
$m\approx\sqrt{2\lambda}a$.

\section{Perturbation scheme}

\subsection{Potential and motion}

\ \ An expansion parameter $g$ is introduced so that we formally write%
\begin{equation}
V(\varphi)=V_{0}(\varphi)+gV_{1}(\varphi) \label{1}%
\end{equation}

where $g$ is an expansion, or control, parameter with $0\leq g\leq1$. (In
computing final corrections, we take the limit $g\rightarrow1$.) When $g=0$
the potential is the unperturbed potential $V_{0}$ and when $g=1$ then $V$ is
the full potential $V=V_{0}+V_{1}$. To compute the set of corrections
$\{\varphi_{n}(x)\}$, $n=1,2,3,...$, to the unperturbed solution $\varphi
_{0}(x)$, a final setting $g=1$ is chosen, but to obtain a set of equations
describing the various orders of corrections $\{\varphi_{n}(x)\}$ the value of
$g$ is temporarily left arbitrary with $g\in\lbrack0,1]$.

\bigskip

\ \ The Lagrangian for the (real) scalar field is%
\begin{equation}
\mathcal{L}=\frac{1}{2}\partial^{\mu}\varphi\partial_{\mu}\varphi-V(\varphi)
\label{2}%
\end{equation}

and units are chosen for which $\hbar=c=1$. The metric is mostly negative with
diag $\eta_{\mu\nu}=(+,-,-,-)$. Furthermore, we define the function
$F(\varphi)$ as the derivative of $V(\varphi)$:
\begin{equation}
F(\varphi)=F_{0}(\varphi)+gF_{1}(\varphi)\equiv\frac{\partial V(\varphi
)}{\partial\varphi}=V^{\prime}(\varphi)=V_{0}^{\prime}(\varphi)+gV_{1}%
^{\prime}(\varphi) \label{3}%
\end{equation}

with the prime denoting differentiation with respect to the argument of the
function, i.e., $F_{0}(\varphi)=V_{0}^{\prime}(\varphi)=dV_{0}(\varphi
)/d\varphi$, $F_{1}(\varphi)=V_{1}^{\prime}(\varphi)$,\ $F^{\prime}%
(\varphi)=V^{\prime\prime}(\varphi)$, etc. The quantity $F(\varphi_{0})$
denotes $F(\varphi)$ evaluated at $\varphi=\varphi_{0}$:
\begin{equation}
F(\varphi_{0})=V^{\prime}(\varphi_{0})=\frac{\partial V(\varphi)}%
{\partial\varphi}\Big|_{\varphi=\varphi_{0}} \label{4}%
\end{equation}

\bigskip

\ \ The equation of motion that follows from $\mathcal{L}$ is%
\begin{equation}
\square\varphi+F(\varphi)=0 \label{5}%
\end{equation}

where $\square=\partial_{t}^{2}-\nabla^{2}$.

\subsection{Expansion scheme}

\ \ As with ordinary Rayleigh-Schr\"{o}dinger perturbation theory in quantum
mechanics, when the perturbing potential $V_{1}$ is modulated by the expansion
parameter $g$, as in (\ref{1}), the field $\varphi$ also becomes dependent
upon $g$ until a particular final setting for $g$ is chosen (say, $g=1$), and
we therefore have $\varphi(x,g)=\varphi_{0}(x)+\delta\varphi(x,g)$ where
$\varphi_{0}(x)$ satisfies the unperturbed equation of motion%
\begin{equation}
\square\varphi_{0}(x)+F_{0}(\varphi_{0})=0 \label{6}%
\end{equation}

with $gV_{1}\rightarrow0$ in (\ref{1}), and it is assumed that $\varphi_{0}$
dominates $\delta\varphi$, $|\delta\varphi|\ll|\varphi_{0}|$ (with the
possible exception where $\varphi_{0}\approx0$, but $\delta\varphi$ is assumed
to remain \textquotedblleft small\textquotedblright\ in some well defined sense).

\bigskip

\ \ We expand $\varphi(x,g)$ in powers of $g$,%
\begin{equation}%
\begin{array}
[c]{ll}%
\varphi(x,g) & =\sum_{n=0}^{\infty}g^{n}\varphi_{n}(x)=\varphi_{0}%
(x)+\delta\varphi(x,g)\medskip\\
\delta\varphi(x,g) & =\sum_{n=1}^{\infty}g^{n}\varphi_{n}(x)=g\varphi
_{1}(x)+g^{2}\varphi_{2}(x)+g^{3}\varphi_{3}(x)+\cdot\cdot\cdot
\end{array}
\label{7}%
\end{equation}

and $V(\varphi)$ and $F(\varphi)=V^{\prime}(\varphi)$ are Taylor expanded
about the zeroth order solution $\varphi_{0}$:%
\begin{equation}
F(\varphi)=F(\varphi_{0})+F^{\prime}(\varphi_{0})(\delta\varphi)+\frac{1}%
{2!}F^{\prime\prime}(\varphi_{0})(\delta\varphi)^{2}+\cdot\cdot\cdot\label{8}%
\end{equation}

Noting that $F(\varphi)=F_{0}(\varphi)+gF_{1}(\varphi)=V_{0}^{\prime}%
(\varphi)+gV_{1}^{\prime}(\varphi)$, the expansion for $F(\varphi)$ takes the
form%
\begin{equation}%
\begin{array}
[c]{ll}%
F(\varphi) & =F_{0}(\varphi)+gF_{1}(\varphi)\smallskip\\
& =\left[  F_{0}(\varphi_{0})+F_{0}^{\prime}(\varphi_{0})(\delta\varphi
)+\frac{1}{2!}F_{0}^{\prime\prime}(\varphi_{0})(\delta\varphi)^{2}+\cdot
\cdot\cdot\right]  \smallskip\\
& +g\left[  F_{1}(\varphi_{0})+F_{1}^{\prime}(\varphi_{0})(\delta
\varphi)+\frac{1}{2!}F_{1}^{\prime\prime}(\varphi_{0})(\delta\varphi
)^{2}+\cdot\cdot\cdot\right]
\end{array}
\label{9}%
\end{equation}

The full system, given by (\ref{5}) can be rewritten with the help of
(\ref{7}) and (\ref{9}) as%
\begin{equation}%
\begin{array}
[c]{ll}
& \square(\varphi_{0}+\delta\varphi)+\left[  F_{0}(\varphi_{0})+F_{0}^{\prime
}(\varphi_{0})(\delta\varphi)+\frac{1}{2!}F_{0}^{\prime\prime}(\varphi
_{0})(\delta\varphi)^{2}+\cdot\cdot\cdot\right]  \medskip\\
& +g\left[  F_{1}(\varphi_{0})+F_{1}^{\prime}(\varphi_{0})(\delta
\varphi)+\frac{1}{2!}F_{1}^{\prime\prime}(\varphi_{0})(\delta\varphi
)^{2}+\cdot\cdot\cdot\right]  =0
\end{array}
\label{10}%
\end{equation}

Using (\ref{7}) for the expansion for $\delta\varphi$, and keeping only up to
$O(g^{3})$ terms, we have, approximately,%
\begin{equation}%
\begin{array}
[c]{ll}%
\square(\varphi_{0}+g\varphi_{1}+g^{2}\varphi_{2}+g^{3}\varphi_{3}%
)\medskip+\left[  F_{0}(\varphi_{0})+F_{0}^{\prime}(\varphi_{0})(g\varphi
_{1}+g^{2}\varphi_{2}+g^{3}\varphi_{3})\right]  & \\
+\left[  gF_{1}(\varphi_{0})+F_{1}^{\prime}(\varphi_{0})(g^{2}\varphi
_{1}+g^{3}\varphi_{2})+\frac{1}{2!}F_{1}^{\prime\prime}(\varphi_{0}%
)(g^{3}\varphi_{1}^{2})\right]  =0 &
\end{array}
\label{11}%
\end{equation}

The various $g^{n}$ terms can be collected to give a set of equations for the
$\varphi_{n}$:%
\begin{equation}%
\begin{array}
[c]{ll}%
g^{0}: & \square\varphi_{0}+F_{0}(\varphi_{0})=0\\
g^{1}: & \square\varphi_{1}+F_{0}^{\prime}(\varphi_{0})\varphi_{1}%
+F_{1}(\varphi_{0})=0\\
g^{2}: & \square\varphi_{2}+F_{0}^{\prime}(\varphi_{0})\varphi_{2}%
+F_{1}^{\prime}(\varphi_{0})\varphi_{1}+\frac{1}{2}F_{0}^{\prime\prime
}(\varphi_{0})\varphi_{1}^{2}=0\\
g^{3}: & \square\varphi_{3}+F_{0}^{\prime}(\varphi_{0})\varphi_{3}%
+F_{1}^{\prime}(\varphi_{0})\varphi_{2}+\frac{1}{2}F_{1}^{\prime\prime
}(\varphi_{0})\varphi_{1}^{2}+F_{0}^{\prime\prime}(\varphi_{0})\varphi
_{1}\varphi_{2}=0
\end{array}
\label{12}%
\end{equation}

\section{Kink and Domain Wall Potential}

\subsection{Zeroth order solution for the unperturbed system}

\ \ We now consider perturbations to time independent topological defects that
depend upon a single coordinate $x$. In particular, attention is given to one
dimensional (1D) kinks and three dimensional (3D) planar domain walls. These
configurations are described by (\ref{2}) and (\ref{5}) with a $\varphi^{4}$
double well potential,%
\begin{equation}
V_{0}(\varphi)=\frac{\lambda}{4}(\varphi^{2}-a^{2})^{2},\ \ \ \ \ F_{0}%
(\varphi)=\lambda\varphi(\varphi^{2}-a^{2}) \label{13}%
\end{equation}

The unperturbed system, with $\varphi=\varphi_{0}$, satisfies $\square
\varphi_{0}+F_{0}(\varphi_{0})=0$ and admits a $\varphi^{4}$ kink/wall
solution%
\begin{equation}
\varphi_{0}(x)=a\tanh(kx)=a\tanh\left(  \frac{x}{w}\right)  =a\tanh\left(
\frac{x}{2\delta}\right)  \label{14}%
\end{equation}

(The antikink/antiwall solution is given by $\varphi_{\bar{K}}(x)=-\varphi
_{K}(x)=-\varphi_{0}(x)$.) The parameters $k$, $w$, and $\delta$ are given by%
\begin{equation}
k=\frac{1}{w}=\frac{1}{2\delta}=\sqrt{\frac{\lambda a^{2}}{2}},\ \ \ w=\sqrt
{\frac{2}{\lambda a^{2}}},\ \ \ \delta=\frac{1}{\sqrt{2\lambda a^{2}}}
\label{15}%
\end{equation}

where $w$ is a \textquotedblleft width\textquotedblright\ parameter for the
kink/wall and $\delta$ is a \textquotedblleft half-width\textquotedblright%
\ parameter. The field $\varphi_{0}$ interpolates between the vacuum values
$\varphi_{0}=\pm a$ where $V_{0}(\pm a)=0$ with $V_{0}(0)=\frac{1}{4}\lambda
a^{4}$.

\subsection{The slab approximation}

\ \ The stress-energy of a planar $\varphi^{4}$ domain wall is
\cite{VSbook,KTbook}
\begin{equation}
T_{\mu\nu}=\partial_{\mu}\varphi\partial_{\nu}\varphi-g_{\mu\nu}%
\mathcal{L}\mathbf{,}\ \ \ T_{\nu}^{\mu}=f(x)\text{\ diag }%
(1,0,1,1),\ \ \ f(x)=\frac{1}{2}\lambda a^{4}\text{sech}^{4}\left(  \frac
{x}{w}\right)  \label{16}%
\end{equation}

with an energy density $T_{00}=f(x)=\frac{1}{2}\lambda a^{4}$sech$^{4}\left(
\frac{x}{w}\right)  $. The surface energy density (energy/unit area) of a
domain wall is \cite{VSbook}%
\begin{equation}
\sigma=\frac{1}{A}\int T_{00}d^{3}x=\int T_{00}dx=\frac{2}{3}\sqrt{2\lambda
}a^{3} \label{17}%
\end{equation}

The energy density $T_{00}$ of a wall is concentrated within the wall's core,
which is centered at $x=0$. The energy density $T_{00}$ peaks at $x=0$ and
then rapidly falls off to zero outside of the wall. It is therefore useful,
making calculations tractable, to employ a \textquotedblleft\textit{slab
approximation}\textquotedblright\cite{JM95}\ for the solution $\varphi_{0}(x)$
wherein $\varphi_{0}(x)$ is taken to be zero inside a slab of
\textquotedblleft\textit{effective thickness}\textquotedblright\ $W=2\Delta,$
where $\Delta$ is considered to be an \textquotedblleft\textit{effective half
width}\textquotedblright\ of the wall (rather than the \textquotedblleft half
width\textquotedblright\ parameter $\delta$) enclosing most of the wall's
energy. (Note that if we imposed boundaries to be located at $x=\pm\delta$,
then at a boundary $kx=\pm k\delta=\pm\frac{1}{2}$ which would give
$\varphi_{0}/a\sim\tanh(\pm k\delta)=\pm\tanh(\frac{1}{2})\sim\pm\frac{1}{2}$,
and $\varphi_{0}$ would fall short of its vacuum values for which $\varphi
_{0}/a\sim\pm1$.) The base solution $\varphi_{0}(x)$ assumes its asymptotic
values of $\pm a$ outside of the slab. The effective half thickness $\Delta$
is to be determined by the application of boundary conditions at the edges of
the slab. The wall's energy is therefore envisioned as being in the form of a
slab of thickness $W=2\Delta$, centered at $x=0$, with $\varphi\approx0$
inside the slab and $\varphi_{0}\approx\pm a$ outside.

\bigskip

\ \ For an energy scale with $\sqrt{\lambda}a$ $\sim O($GeV$)$, we expect the
slab thickness to be quite small, with $\Delta\sim O(\delta)\sim O($%
GeV$^{-1})$, i.e., $\Delta\sim O(.2$ fm$)$. (The Nambu action for a domain
wall approximates the wall as a sheet of zero thickness. See, for example,
\cite{note} for a description of the Nambu action for a domain wall, and
\cite{Gregory90} for thickness corrections to the Nambu action.)

\bigskip

\ \ Specifically, within the slab approximation, we make the approximations
$\tanh(kx)\approx0$ for $kx\in(-k\Delta,k\Delta)$ and $|\tanh(kx)|\approx1$
for $k|x|>k\Delta$, i.e.,
\begin{subequations}
\label{18}%
\begin{align}
\tanh(kx)  &  \approx\left\{
\begin{array}
[c]{lll}%
0, & \ \ x\in(-\Delta,\Delta) & \text{(inside wall)}\\
-1, & \ \ x<-\Delta & \text{(outside wall)}\\
+1, & \ \ x>\Delta & \text{(outside wall)}%
\end{array}
\right\} \label{18a}\\
|\tanh(kx)|  &  \approx\left\{
\begin{array}
[c]{lll}%
0, & \ \ x\in(-\Delta,\Delta) & \text{(inside wall)}\\
1, & \ \ |x|>\Delta & \text{(outside wall)}%
\end{array}
\right\}  \label{18b}%
\end{align}
\newline

\ \ Constructing correction solutions for $\varphi_{n}(x)$ using the slab
approximation is then somewhat analogous to finding quantum mechanical
solutions $\psi(x)$ for a finite square well potential, where boundary
conditions include continuity of the wave function $\psi(x)$ and its first
derivative $\psi^{\prime}(x)$ at the discontinuous boundaries of the well,
along with appropriate asymptotic behaviors. Likewise, at least for the
complimentary solutions $\varphi_{c}$ of the homogeneous DEs (where
$F_{1}\rightarrow0$), we assume continuity of $\varphi_{c}(x)$ and
$\varphi_{c}^{\prime}(x)$ at $x=\pm\Delta$, along with appropriate asymptotic
behaviors for $\varphi$ as $x\rightarrow\pm\infty$ in order to obtain quick
estimates for the effective width $\Delta$.

\subsection{First order corrections}

\ \ The first order correction $\varphi_{1}(x)$ (the $g^{1}$ equation in
(\ref{12})), i.e., $\square\varphi_{1}+F_{0}^{\prime}(\varphi_{0})\varphi
_{1}+F_{1}(\varphi_{0})=0$, reads as%
\end{subequations}
\begin{equation}
\partial_{x}^{2}\varphi_{1}-F_{0}^{\prime}(\varphi_{0})\varphi_{1}%
=F_{1}(\varphi_{0}) \label{19}%
\end{equation}

where $\partial_{x}^{2}=\partial^{2}/\partial x^{2}$ and $F_{0}^{\prime
}(\varphi_{0})=(\partial F_{0}(\varphi)/\partial\varphi)|_{\varphi_{0}}$. The
inhomogeneous term for this second order ordinary differential equation (DE)
is $F_{1}(\varphi_{0})=V_{1}^{\prime}(\varphi)|_{\varphi_{0}}$, which is left
unspecified for the time being. The solution $\varphi_{1}(x)$ consists of a
\textquotedblleft complimentary function\textquotedblright\ $\varphi
_{1,c}(x)\equiv\psi_{c}(x)$ along with a \textquotedblleft particular
solution\textquotedblright\ $\psi_{p}(x),$ so that $\varphi_{1}(x)=\psi
_{c}(x)+\psi_{p}(x)$. It is assumed that, in principle, the DE (\ref{19})
possesses, in reality, a sufficiently smooth, continuous inhomogeneous
function $F_{1}(\varphi_{0})$ for the smooth function $\varphi_{0}\sim
\tanh(kx)$, along with a smooth, continuous homogeneous solution $\psi_{c}$.
However, for a given function $F_{1}(\varphi)$, an implementation of the slab
approximation for $\varphi_{0}$ and $F(\varphi_{0})$ can, in general, destroy
the global continuity of $F_{1}$ within this approximation, requiring instead
a piecewise continuity of $F_{1}(\varphi_{0})$.

\bigskip

\ \ The base solution is $\varphi_{0}(x)=a\tanh(kx)$ and $F_{0}(\varphi)$ is
given by (\ref{13}) with $F_{0}^{\prime}(\varphi)=\lambda(3\varphi^{2}-a^{2}%
)$. We therefore have%
\begin{equation}
F_{0}^{\prime}(\varphi_{0})=\lambda a^{2}\left[  3\tanh^{2}(kx)-1\right]
=2k^{2}\left[  3\tanh^{2}(kx)-1\right]  \label{20}%
\end{equation}

with $2k^{2}=\lambda a^{2}$. The DE (\ref{19}) therefore becomes%
\begin{equation}
\partial_{x}^{2}\varphi_{1}-2k^{2}\left[  3\tanh^{2}(kx)-1\right]  \varphi
_{1}=F_{1}(\varphi_{0}) \label{21}%
\end{equation}

Setting $F_{1}=0$ gives the homogeneous DE for the complimentary function
$\varphi_{1,c}=\psi_{c}$:%
\begin{equation}
\partial_{x}^{2}\psi_{c}-2k^{2}\left[  3\tanh^{2}(kx)-1\right]  \psi_{c}=0
\label{22}%
\end{equation}

Since $\varphi_{0}(x)=a\tanh(kx)$ is a smooth, continuous function of $x$, we
expect that for a smooth, continuous function $F_{1}(\varphi_{0}(x))$ there
exists a smooth, continuous solution $\varphi_{1}(x)$ that is compatible with
boundary conditions. However, if we implement the slab approximation,
discontinuities are introduced which may show up in the functions $\psi
_{p}(x)$ and therefore $\varphi_{1}(x)$. These discontinuities may then
require the consideration of piecewise continuous functions for $\psi_{p}$ and
$\varphi_{1}$.

\bigskip

\ \ Eq.(\ref{22}) is difficult to solve in terms of simple elementary
functions. However, it is interesting to comment that it resembles a
Schr\"{o}dinger equation for a potential $U(x)=3k^{2}\tanh^{2}(kx)$, i.e.,
\begin{equation}
-\frac{1}{2}\frac{\partial^{2}\psi_{c}}{\partial x^{2}}+3k^{2}\tanh
^{2}(kx)\psi_{c}=k^{2}\psi_{c} \label{23}%
\end{equation}

with an \textquotedblleft energy\textquotedblright\ eigenvalue $E=k^{2}$ (with
a mass $m$ set equal to unity). If the $\tanh^{2}$ potential is approximated
by a finite square well of height $U_{0}=3E$ and width $W=2\Delta$, then there
exist eigenstates of the system which are bound states, and the eigenfunctions
are subjected to boundary conditions involving continuity of $\psi_{c}(x)$ and
$\psi_{c}^{\prime}(x)$ at the edges of the potential well, i.e., at
$x=\pm\Delta$. In addition, $\psi_{c}$ must remain finite at $x=\pm\infty$.
(For the field theory case, we do not require that the function $\psi_{c}$
normalize to unity.)

\bigskip

\ \ We now proceed to invoke the slab approximation for the $\tanh^{2}(kx)$
function in (\ref{22}), in which case, by (\ref{18}), (\ref{22}) is
represented by the slab DEs
\begin{subequations}
\label{24}%
\begin{align}
\psi_{c}^{\prime\prime}(x)+2k^{2}\psi_{c}(x)  &  =0,\ \ x\in(-\Delta
,\Delta)\label{24a}\\
\psi_{c}^{\prime\prime}(x)-4k^{2}\psi_{c}(x)  &  =0,\ \ |x|>\Delta\label{24b}%
\end{align}

The differential operator $\partial^{2}/\partial x^{2}$ and the slab potential
for $\tanh^{2}(kx)$ are even in $x$, so that the solutions $\psi_{c}(x)$ must
have definite parity with $\psi_{c}(x)=\pm\psi_{c}(-x)$. Also, it is required
that $\psi_{c}$ be finite at $x=\pm\infty$. The solution set for (\ref{24}) is
therefore given by%
\end{subequations}
\begin{equation}
\psi_{c}(x)=\left\{
\begin{array}
[c]{cc}%
\left\{
\begin{array}
[c]{cc}%
\psi_{c}^{(even)}=D\cos(\sqrt{2}kx) & \\
\psi_{c}^{(odd)}=C\sin(\sqrt{2}kx) &
\end{array}
\right\}  , & x\in(-\Delta,\Delta)\\
Ae^{2kx}\ , & \ x<-\Delta\\
\ Be^{-2kx}\ , & x>\Delta
\end{array}
\right\}  \label{25}%
\end{equation}

with%
\begin{equation}
\psi_{c}^{\prime}(x)=\left\{
\begin{array}
[c]{cc}%
\left\{
\begin{array}
[c]{cc}%
\psi_{c}^{\prime(even)}=-\sqrt{2}kD\sin(\sqrt{2}kx) & \\
\psi_{c}^{\prime(odd)}=\sqrt{2}kC\cos(\sqrt{2}kx) &
\end{array}
\right\}  , & x\in(-\Delta,\Delta)\\
2kAe^{2kx}\ , & x<-\Delta\\
-2kBe^{-2kx}\ , & x>\Delta
\end{array}
\right\}  \label{26}%
\end{equation}

\ Upon imposing boundary conditions that $\psi_{c}$ and $\psi_{c}^{\prime}$ be
continuous at $x=\pm\Delta$, we get the following results for the even and odd
solutions for a determination of $k\Delta$.

\bigskip

\ \ \textbf{Even solutions:} $\psi_{c}(x)=\psi_{c}(-x)$,\ \ ($A=B$): (See
(\ref{25}).)%
\begin{equation}%
\begin{array}
[c]{ll}%
D\cos(\sqrt{2}k\Delta)=Ae^{-2k\Delta} & \\
D\sin(\sqrt{2}k\Delta)=\sqrt{2}Ae^{-2k\Delta} & \\
\implies\tan(\sqrt{2}k\Delta)=\sqrt{2} &
\end{array}
\label{27}%
\end{equation}

\ \ From (\ref{27}) we have $\tan(\sqrt{2}k\Delta)=\sqrt{2}$ which is solved
by $k\Delta_{n}^{(even)}=\frac{1}{\sqrt{2}}\left[  \tan^{-1}(\sqrt{2}%
)+n\pi\right]  ,\ \ n\in\mathbb{Z}$ $\ (n=...,-2,-1,0,1,2,...)$. For values of
$k\Delta>0$ we require $n=0,1,2,...$, with $k\Delta_{n=0}^{(even)}=\frac
{1}{\sqrt{2}}\tan^{-1}(\sqrt{2})\approx.675$. Therefore, the physical solution
is given by $k\Delta_{n}^{(even)}=\frac{1}{\sqrt{2}}\left[  \tan^{-1}(\sqrt
{2})+n\pi\right]  ,\ n\geq0$ . This allows the determination of $k\Delta
_{n}^{(even)}\geq0$ for a given value of $k=\sqrt{\lambda a^{2}/2}$,
satisfying the condition $\tan(\sqrt{2}k\Delta)=\sqrt{2}$ for \textbf{even}
solutions $\psi_{c}(x)$.

\bigskip

\ \ \textbf{Odd solutions:} $\psi_{c}(x)=-\psi_{c}(-x)$,\ \ ( $A=-B$): (See
(\ref{25}).)%
\begin{equation}%
\begin{array}
[c]{ll}%
C\sin(\sqrt{2}k\Delta)=Be^{-2k\Delta} & \\
C\cos(\sqrt{2}k\Delta)=-\sqrt{2}Be^{-2k\Delta} & \\
\implies\tan(\sqrt{2}k\Delta)=-1/\sqrt{2} &
\end{array}
\label{28}%
\end{equation}

\ \ From (\ref{28}) we have $\tan(\sqrt{2}k\Delta)=-\frac{1}{\sqrt{2}}$ which
is solved by $k\Delta_{n}^{(odd)}=\frac{1}{\sqrt{2}}\left[  -\tan^{-1}%
(\frac{1}{\sqrt{2}})+n\pi\right]  =-\frac{1}{\sqrt{2}}\left[  \tan^{-1}%
(\frac{1}{\sqrt{2}})-n\pi\right]  ,$ $n\in\mathbb{Z}$. For $k\Delta
_{n}^{(odd)}>0$ we require $n\geq1$. Therefore, $k\Delta_{n}^{(odd)}\geq
k\Delta_{n=1}^{(odd)}\approx1.78.$ This allows the determination of
$k\Delta_{n}^{(odd)}\geq0$ for a given value of $k=\sqrt{\lambda a^{2}/2}$,
satisfying the condition $\tan(\sqrt{2}k\Delta)=-1/\sqrt{2}$ for \textbf{odd}
solutions $\psi_{c}(x)$.

\bigskip

\ \ In summary, it is found that $k\Delta_{n}=(\sqrt{\lambda a^{2}/2}%
\ )\Delta_{n}$ takes values\footnote{$\frac{1}{\sqrt{2}}\tan^{-1}(\sqrt
{2})\approx.676,\ \ $and$\ -\frac{1}{\sqrt{2}}\tan^{-1}(\frac{1}{\sqrt{2}%
})\approx-.435$}%
\begin{equation}
k\Delta_{n}=\left\{
\begin{array}
[c]{ll}%
k\Delta_{n}^{(even)}=\frac{1}{\sqrt{2}}\left[  \tan^{-1}(\sqrt{2}%
)+n\pi\right]  \gtrsim.675,\ (n=0,1,2,...) & \text{(even solutions)}\\
k\Delta_{n}^{(odd)}=\frac{1}{\sqrt{2}}\left[  -\tan^{-1}(\frac{1}{\sqrt{2}%
})+n\pi\right]  \gtrsim1.78,\ (n=1,2,...) & \text{(odd solutions)}%
\end{array}
\right\}  \label{29}%
\end{equation}
allowing a rough approximation for a determination of an effective slab width
$W_{n}=2\Delta_{n}$ for even or odd solutions $\psi_{c}(x)$. (It is noted
later that a more exact determination of $k\Delta$ with an inclusion of the
inhomogeneous term $F_{1}(\varphi_{0})$ and the particular solution $\psi
_{p}(x)$ requires solving a much more complicated algebra problem. However, a
qualitative justification for the approximate accuracy of the estimates
obtained above is provided in an illustration.)

\bigskip

\ \ These approximate values of $k\Delta$ emerging from the complimentary
functions $\psi_{c}(x)$ illustrate the following salient features.

\bigskip

\ \ \ (i) The \textit{effective width} $\Delta$ is \textit{quantized} in order
to admit either even or odd solutions $\psi_{c}$ to (\ref{24}). This is
analogous to the quantum mechanical result for a finite square well where the
\textit{energy eigenvalues} for even and odd wavefunctions are quantized for a
given well width.

\bigskip

\ \ \ (ii) There is always an even solution $\psi_{c}$ for $k\Delta_{n}$ with
$n=0$, but there are no odd solutions $\psi_{c}$ for $k\Delta_{n}$ unless
$n\geq1$.

\bigskip

\ \ (iii) It is found that $k\Delta_{n}^{(even)}\geq k\Delta_{n=0}%
^{(even)}\approx.675,\ (n\geq0)$ and $k\Delta_{n}^{(odd)}\geq k\Delta
_{n=1}^{(odd)}\approx1.78,\ \ (n\geq1)$. These can be compared to the value
$k\delta=.5$ for the \textquotedblleft half width\textquotedblright\ parameter
$\delta$, showing that $\Delta\gtrsim\delta$, with $\Delta\sim O(\delta)$.

\bigskip

\ \ (iv) The fact that $k\Delta_{\text{min}}^{(odd)}>k\Delta_{\text{min}%
}^{(even)}$ can be understood in analogy to the solutions for the square well
problem: in order to accommodate the boundary conditions in (\ref{28}), one
must be able to fit at least $\frac{1}{4}$ of a wavelength between $x=0$ and
$x=\Delta$. In other words, $\theta_{\text{min}}\equiv k\Delta_{\text{min}%
}^{(odd)}\gtrsim\left(  \frac{2\pi}{\lambda}\right)  (\frac{\lambda}{4}%
)=\frac{\pi}{2}=\allowbreak1.\,\allowbreak57$, which is near the value of
$k\Delta_{\text{min}}^{(odd)}\approx1.78$ determined above.

\section{Illustration}

\subsection{The potential}

\ \ As an example of the application of the perturbation method, along with
the use of the slab approximation, we consider a perturbation given by%
\begin{equation}
V_{1}(\varphi)=\frac{1}{3}\mu\varphi^{3},\ \ \ V_{1}^{\prime}(\varphi
)=F_{1}(\varphi)=\mu\varphi^{2} \label{30}%
\end{equation}

where $\mu$ is a small constant with dimensions of mass, and we take $\mu>0$
for definiteness. It is endeavored to determine the first order correction
$\varphi_{1}(x)$ to the unperturbed solution $\varphi_{0}(x)$. (We now adopt
the setting $g=1$.) The total solution is then $\varphi(x)=\varphi
_{0}(x)+\varphi_{1}(x)$. Furthermore, the complimentary functions $\psi
_{c}(x)$ have been determined, given by (\ref{25}).

\bigskip

\ \ The total potential (with $g\rightarrow1$) is%
\begin{equation}
V(\varphi)=V_{0}(\varphi)+V_{1}(\varphi)=\frac{1}{4}\lambda(\varphi^{2}%
-a^{2})^{2}+\frac{1}{3}\mu\varphi^{3} \label{31}%
\end{equation}

and it is assumed that $\mu\ll\lambda a$. The $V_{1}$ term causes the
potential $V$ to become slightly asymmetric, with $V^{+}\equiv V(a)=\frac
{1}{3}\mu a^{3}$ and $V^{-}\equiv V(-a)=-\frac{1}{3}\mu a^{3}$ so that in
vacuum we have, approximately, $\Delta V=V^{+}-V^{-}=\frac{2}{3}\mu a^{3}$.
The locations of the vacuum states are determined by $F(\varphi)=V^{\prime
}(\varphi)=0$, where%
\begin{equation}
F(\varphi)=F_{0}(\varphi)+F_{1}(\varphi)=\varphi\left[  \lambda(\varphi
^{2}-a^{2})+\mu\varphi\right]  \label{32}%
\end{equation}

The condition $F(\varphi)|_{\varphi_{vac}}=0$ yields%
\begin{equation}
\varphi_{vac}=\pm a\left(  1+\frac{\mu^{2}}{4\lambda^{2}a^{2}}\right)
^{1/2}-\left(  \frac{\mu}{2\lambda a}\right)  a\approx\pm a-\left(  \frac{\mu
}{2\lambda a}\right)  a\approx\pm a \label{33}%
\end{equation}

Since $\frac{\mu}{\lambda a}\ll1$, for the sake of convenience we approximate
$\varphi_{vac}\approx\pm a$ for the evaluation of $V(\varphi)$ in the vacuum
states, so that as above, $V^{\pm}=V(\varphi_{vac}=\pm a)=\pm\frac{1}{3}\mu
a^{3}$.

\bigskip

\ \ Now notice that vacuum states located by $\varphi_{vac}=\pm a-\left(
\frac{\mu}{2\lambda a}\right)  a$ of (\ref{33}) can also be obtained by using
the slab approximation. Writing $\varphi_{1}(x)=\psi_{c}(x)+\psi_{p}(x)$ along
with (\ref{24}) for the DEs for $\psi_{c}$, the slab approximation for
$\varphi_{0}(x)$ allows (\ref{21}) to be expressed by the interior and
exterior DEs for the particular solution $\psi_{p}$:%
\begin{equation}
\left\{
\begin{array}
[c]{ll}%
\psi_{p}^{\prime\prime}+2k^{2}\psi_{p}=F_{1}(\varphi_{0})=0, & x\in
(-\Delta,\Delta),\ \ \varphi_{0}=0\\
\psi_{p}^{\prime\prime}-4k^{2}\psi_{p}=F_{1}(\varphi_{0})=\mu a^{2}, &
|x|>\Delta,\ \ \varphi_{0}^{2}=a^{2}%
\end{array}
\right\}  \label{33a}%
\end{equation}

Since the inhomogeneous terms are constants, we take $\psi_{p}=$ const, in
which case $\psi_{p}^{\prime\prime}=0$. Eq.(\ref{33a}) is then solved by%
\begin{equation}
\psi_{p}=\left\{
\begin{array}
[c]{ccc}%
0, & x\in(-\Delta,\Delta), & \varphi_{0}=0\\
-\frac{1}{2}\left(  \dfrac{\mu}{\lambda a}\right)  a, & |x|>\Delta, &
|\varphi_{0}|=a
\end{array}
\right\}  \label{33b}%
\end{equation}

The first order correction is then given by%
\begin{equation}
\varphi_{1}=\psi_{c}+\psi_{p}=\left\{
\begin{array}
[c]{ll}%
\left\{
\begin{array}
[c]{c}%
D\cos(\sqrt{2}kx)\\
C\sin(\sqrt{2}kx)
\end{array}
\right\}  , & x\in(-\Delta,\Delta),\ \ \varphi_{0}=0\\
Ae^{2kx}-\frac{1}{2}\left(  \frac{\mu}{\lambda a}\right)  a, & x<-\Delta
,\ \ \varphi_{0}=-a\\
Be^{-2kx}-\frac{1}{2}\left(  \frac{\mu}{\lambda a}\right)  a, & x>\Delta
,\ \ \varphi_{0}=+a
\end{array}
\right\}  \label{33c}%
\end{equation}

\ and the full solution $\varphi=\varphi_{0}+\varphi_{1}$ is%
\begin{equation}
\varphi=\varphi_{0}+\varphi_{1}=\left\{
\begin{array}
[c]{ll}%
\left\{
\begin{array}
[c]{c}%
D\cos(\sqrt{2}kx)\\
C\sin(\sqrt{2}kx)
\end{array}
\right\}  , & x\in(-\Delta,\Delta),\ \ \varphi_{0}=0\\
Ae^{2kx}-a-\frac{1}{2}\left(  \frac{\mu}{\lambda a}\right)  a, &
x<-\Delta,\ \ \varphi_{0}=-a\\
Be^{-2kx}+a-\frac{1}{2}\left(  \frac{\mu}{\lambda a}\right)  a, &
x>\Delta,\ \ \varphi_{0}=+a
\end{array}
\right\}  \label{33d}%
\end{equation}

\ \ The constants are left undetermined here, but $\psi_{c}(x)$ and $\psi
_{c}^{\prime}(x)$ have previously been subjected to a matching of interior and
exterior solutions at the boundaries $x=\pm\Delta$ (for the case $\mu=0$)
which, along with an account of solution parity, allows for a reduction of the
constants (say $B,C,$ and $D$ in terms of $A$), as in the quantum mechanical
case of a finite square well. (See, e.g., (\ref{27})-(\ref{29}) \ for the
cases $B=\pm A$. The smallness of $\psi_{p}$ is expected to have negligible
impact on the full solution $\varphi$, and therefore little impact upon the
determination of effective slab width $\Delta$.)

\bigskip

\ \ Matching the full slab approximated solutions for $\varphi_{1}(x)=\psi
_{c}(x)+\psi_{p}$ (with $\psi_{p}\neq0$) at the boundaries $x=\pm\Delta$
presents a more convoluted algebraic problem than for the $\mu=0$, $\psi
_{p}=0$ case. Since we do not expect a small perturbation $V_{1}$ to make a
significant change to the unperturbed solution $\varphi_{0}$, we content
ourselves with the estimates given in (\ref{29}) as approximate
representations of the effective slab widths using the slab approximation, as
these estimates are independent of the constants $A,B,C,$ and $D$ appearing in
the complimentary function part $\psi_{c}$ of the correction $\varphi_{1}$.

\subsection{Vacuum bubbles}

\ \ The stress-energy of the domain wall, from (\ref{16}), is%
\begin{equation}
T_{\mu\nu}=\partial_{\mu}\varphi\partial_{\nu}\varphi+\eta_{\mu\nu}\left[
\frac{1}{2}(\partial_{x}\varphi)^{2}+V(\varphi)\right]  \label{34}%
\end{equation}

Multiplication of the equation of motion $\partial_{x}^{2}\varphi
=\partial_{\varphi}V$ by $\partial_{x}\varphi$ and an integration leads to%
\begin{equation}
\frac{1}{2}(\partial_{x}\varphi)^{2}=V(\varphi)+K \label{35}%
\end{equation}

where $K$ is a constant of integration. The stress-energy tensor is then%
\begin{equation}
T_{\mu\nu}=\partial_{\mu}\varphi\partial_{\nu}\varphi+\eta_{\mu\nu}\left[
2V(\varphi)+K\right]  \label{36}%
\end{equation}

This gives, by (\ref{35}),%
\begin{equation}
T_{00}=2V+K,\ \ \ T_{xx}=K,\ \ \ T_{yy}=T_{zz}=-T_{00}=-(2V+K) \label{37}%
\end{equation}

However, due to the asymmetry of the potential, as $x\rightarrow\pm\infty$ and
$\varphi\rightarrow\pm a$, then the potential takes asymptotic values
$V(\varphi)\rightarrow V^{\pm}$ with $V^{+}=+\frac{1}{3}\mu a^{3}$ and
$V^{-}=-\frac{1}{3}\mu a^{3}$. From (\ref{35}) we have positive and negative
domains $(\partial_{x}\varphi)_{\pm}^{2}$, for which there are two different
integration constants $K^{\pm}$, and therefore (\ref{35}) gets modified to%
\begin{equation}
\frac{1}{2}(\partial_{x}\varphi)_{\pm}^{2}=V(\varphi)+K^{\pm} \label{38}%
\end{equation}

Since $\partial_{x}\varphi\rightarrow0$ in vacuum, then $V^{\pm}=-K^{\pm}%
=\pm\frac{1}{3}\mu a^{3}$, and the asymptotic stress-energy splits into two
parts, $T_{\mu\nu}(\pm\infty)\rightarrow T_{\mu\nu}^{\pm}$. We then have
(\ref{36}) taking the form%
\begin{equation}
T_{\mu\nu}=\partial_{\mu}\varphi\partial_{\nu}\varphi+\eta_{\mu\nu}\left[
2V(\varphi)+K^{\pm}\right]  =\partial_{\mu}\varphi\partial_{\nu}\varphi
+\eta_{\mu\nu}\left[  2V(\varphi)-V^{\pm}\right]  \label{39}%
\end{equation}

In vacuum, $\partial_{\mu}\varphi\rightarrow0$, $V(\varphi)\rightarrow V^{\pm
}$, $K^{\pm}=-V^{\pm}$, and asymptotically, as $x\rightarrow\pm\infty$, the
stress-energy becomes%
\begin{equation}
T_{\mu\nu}\rightarrow T_{\mu\nu}^{\pm}=\eta_{\mu\nu}V^{\pm}=\pm\eta_{\mu\nu
}\left(  \tfrac{1}{3}\mu a^{3}\right)  ,\ \ \ \ \ |x|\rightarrow
\infty\label{40}%
\end{equation}

The components of $T_{\mu\nu}^{\pm}$ are
\begin{equation}
T_{00}^{\pm}=\pm V^{\pm}=\pm\tfrac{1}{3}\mu a^{3},\ \ \ T_{xx}^{\pm}%
=T_{yy}^{\pm}=T_{zz}^{\pm}=-T_{00}^{\pm}=\mp V^{\pm}=\mp\tfrac{1}{3}\mu a^{3}
\label{41}%
\end{equation}

This stress-energy $T_{\mu\nu}^{\pm}=T_{\mu\nu}(x=\pm\infty)$ is quickly
acquired outside the core of the domain wall. In the slab approximation we
have, roughly, $T_{\mu\nu}\approx T_{\mu\nu}^{\pm}=\pm\eta_{\mu\nu}\left(
\frac{1}{3}\mu a^{3}\right)  $ for $|x|>\Delta$ just outside of the slab.

\bigskip

\ \ At any rate, effectively $T_{\mu\nu}\approx\pm\eta_{\mu\nu}\left(
\frac{1}{3}\mu a^{3}\right)  $ on the outer edges of the domain wall, and the
tangential stresses $T_{yy}=T_{zz}=-T_{00}$ are higher on one side than the
other, with $|\Delta T_{yy}|=|T_{yy}^{+}-T_{yy}^{-}|\approx\frac{2}{3}\mu
a^{3}$, etc., leading to an instability against bending.\ The bending tends to
occur toward the lower $T_{yy}$, $T_{zz}$ side, i.e., toward the higher
$T_{00}$ side. Because of this instability we conclude that walls collapse
\cite{Vilenkin81}, and that a network of bubbles eventually forms
\cite{Coulson}. A bubble encloses a region of slightly higher energy density
and is surrounded by a region of the lower energy density - the true vacuum.
Such a network formation can occur due to self intersecting trajectories of a
domain wall, and is enhanced by collisions of bending and/or vibrating walls
and antiwalls. Without some efficient stabilization mechanism, the bubbles
collapse with the release of radiation in the form of $\varphi$ boson
particles of mass $m\approx\sqrt{2\lambda}a$.

\section{Summary}

\ \ A Rayleigh-Schr\"{o}dinger type of perturbation scheme is used to study a
self-interacting scalar field with weak perturbations to a potential which
admits known analytic solutions. In particular, the $\varphi^{4}$ double well
potential $V_{0}(\varphi)=\frac{1}{4}\lambda(\varphi^{2}-a^{2})^{2}$ occurring
in models describing 1D domain kinks and 3D domain walls is investigated. The
exact solutions for the unperturbed domain defects, described by $\varphi
_{0}(x)=a\tanh(kx)$, are modified by the perturbing potentials.\-\bigskip

\ This method is illustrated by adding a $V_{1}(\varphi)=\frac{1}{3}\mu
\varphi^{3}$ cubic potential perturbation to the familiar $\varphi^{4}$
quartic kink potential $V_{0}$. A \textquotedblleft slab
approximation\textquotedblright\ is employed and the first order corrections
$\varphi_{1}(x)$ to the unperturbed solution are found, allowing an
approximate representation of the solution $\varphi(x)=\varphi_{0}%
(x)+\varphi_{1}(x)$ for the scalar field theory with potential $V(\varphi
)=V_{0}(\varphi)+V_{1}(\varphi)$. A result is the appearance of an asymmetric
scalar potential $V(\varphi)$ with slightly different, nondegenerate, vacuum
values. Consequently, the domain walls become unstable against bending, with
the subsequent formation of a network of vacuum bubbles. Within the context of
this single scalar field model, the vacuum bubbles collapse, releasing
radiation in the form of $\varphi$ boson particles.

\end{document}